# Medical Image Registration using Unsupervised Deep Neural Network: A Scoping Literature Review

Samaneh Abbasi[a], Meysam Tavakoli[b], Hamid Reza Boveiri[c], Mohammad Amin Mosleh Shirazi[d], Raouf Khayami[c], Hedieh Khorasani[e], Reza Javidan[c], Alireza Mehdizadeh[f]

*[a]Department of Medical Physics and Engineering, School of Medicine, Shiraz University of Medical Sciences, Shiraz, Iran.*
*[b]Department of Radiation Oncology, University of Pittsburgh School of Medicine and UPMC Hillman Cancer Center, Pittsburgh, PA, USA.*
*[c]Department of Computer Engineering and IT, Shiraz University of Technology, Shiraz, Iran.*
*[d]Ionizing and Non-Ionizing Radiation Protection Research Center, School of Paramedical Sciences, Shiraz University of Medical Sciences, Shiraz, Iran. , [e]Department of Medical Informatics, Mashhad University of Medical Science, Mashhad, Iran.*
*[f]Research Center for Neuromodulation and Pain, Shiraz University of Medical Sciences, Shiraz, Iran.*

**Abstract**

In medicine, image registration is vital in image-guided interventions and other clinical applications. However, it is a difficult subject to be addressed which by the advent of machine learning, there have been considerable progress in algorithmic performance has recently been achieved for medical image registration in this area. The implementation of deep neural networks provides an opportunity for some medical applications such as conducting image registration in less time with high accuracy, playing a key role in countering tumors during the operation. The current study presents a comprehensive scoping review on the state-of-the-art literature of medical image registration studies based on unsupervised deep neural networks is conducted, encompassing all the related studies published in this field to this date. Here, we have tried to summarize the latest developments and applications of unsupervised deep learning-based registration methods in the medical field. Fundamental and main concepts, techniques, statistical analysis from different viewpoints, novelties, and future directions are elaborately discussed and conveyed in the current comprehensive scoping review. Besides, this review hopes to help those active readers, who are riveted by this field, achieve deep insight into this exciting field.

**Contents**

## 1. Introduction

In medical imaging, it is sometimes needed to register two images for some purposes such as diagnosis, prognosis, treatment, and follow-up [1]. Image registration is defined as the proper procedure for transforming different images into one coordinate system. In this process, we align multiple images of the same object into a single cumulative amalgamated image, taking into account the rotation, skew, and scale of each separate image. This is what is called image registration, known as image fusion or image matching [2]. An example where image registration can be used is in magnetic resonance imaging (MRI) [3], including $T_1$ and $T_2$ weighted MRI images and computed tomography (CT) [4]. Here, the process of registration is vital procedure in different clinical application such as image-guided [5, 6], motion tracking
[7], segmentation [8, 9], dose accumulation [10, 11, 12], image reconstruction [13, 14]. There are some challenges in the registration process, such as images with different modalities, low-quality, deformation, organ motion, and the noise of interventional images [15].

In the past, image registration was manually done by clinicians in such a way that most of the registration tasks were dependent on the user, i.e. the system was not able to work independently that was a problem; thus, to address this concern, many automatic registration methods have been explored and developed [16]. Conventional registration techniques are optimization procedures based on iteration that need to extract appropriate features, select a similarity measure for the evaluation of registration quality, and decide the transformation model [17]. In these techniques, a pair of images are inputted into the system as fixed and moving images, and then the best alignment can be obtained by constantly sliding the moving image on the fixed image. At first, the connection and relationship between the fixed and moving images are investigated based on a similarity measure. Secondly, an optimization algorithm computes the new transformation parameters. The moving image finds a better correlation with the fixed image in every iteration, and this process continues until no better registration is found. In optimization methods, large calculation processes and parameter sensitivity optimization algorithms, associated with high extraction results [18, 19, 20]; which make the iterative process time-consuming [21]. In this case, a regular deformable image registration needs about 10 min to be run. Beside this, there are different challenges that registration approaches dealing with.

These challenges include, multi-modal registration which is elaborating task to design precise image similarity measures because of the inherent appearance changes between different modalities [2]. Inter-patient registration would be difficult due to the fact that anatomical structures are different across patients. Many studies have been presented to handle the abovementioned issues. Well-known registration algorithms include Elastix [22], optical flow [23, 24], Hammer [25], demons [26], and ANTs [27].

By emerging of deep learning-based approiaches, they have changed the landscape of medical imaging research and obtained the-state-of-art performances in many clinical applications. Thereore, to address above limitation, learning-based techniques and deep neural networks (DNNs), considered robust methods in the community of machine intelligence, can be used for different image processing applications, especially in medical image registration [28]. Recently, various medical image studies have focused on supervised deep leaning-based registration as the main task. A major downside is most of them depend on the ground-truth set, which is considered the primary drawback due to the dependency on the experts for labeling [29, 30]. Therefore, an unsupervised learning is implemented that does not need any ground-truth set [31] and mostly proposes neural networks such as a convolutional neural network (CNN) [32, 33, 34], generative adversarial networks (GAN) [35, 36, 37, 38], and spatial transformation network (STN) [39, 40] that warps images when overlaying with each other. Since there is no ground-truth set in unsupervised approaches, instead, there are some similarity metrics that play as our assessment criteria. These metrics include mutual information (MI) [41, 42, 43, 44], Scale invariant feature transform (SIFT) [45, 46, 47], and local cross-correlation (LCC) [48, 49], that are evaluator criteria. In the same direction, since annotating ground-truth data is another challenge and it is time- and cost-consuming in image registration, the unsupervised method was introduced [50, 51]. Although several review papers on deep learning in medical image registration and analysis have been published [2, 52, 28, 53, 54], to our best of knowlege, there is not any review paper in unsupervised deep learning-based image registration. This scoping review aims to investigate the state-of-the-art literature on medical image registration studies based on unsupervised deep neural network techniques. The evidence for medical image registration indicators based on the unsupervised approach was elaborately synthesized in the current study through a scoping review method [55], given that the principal aims of a scoping review are to summarize and report on research results.

## 2. Reference gathering methodology

In the present study, we emulate a systematic scoping review conducted in 2020 by Arksey and O'Malley. Their approach specifies the criteria for medical image registration based on the unsupervised method. This approach is composed of five main stages in addition to one optional stage: 1) selecting the research question, 2) searching for related studies, 3) refining the studies, 4) tabulating and synthesizing main information, 5) summarizing and reporting the results and 6) validating and verifying the results by the expert panel (optional) [55] that were discussed as follows.

### *2.1. Selecting the research question*

In general, clarity in introducing the structure of the research question is the first step of conducting a scoping review [55, 56]. The main guidelines in prior literature for succession in this step are being welcome in the process of question identification as iterative, clarifying the question, making clear its relationship to the study purpose, and directing the applicable context in conversation with learning from existing states of arts. Here, in this proposed review, the research question is "What are the most common evaluation metrics for medical image registration based on the unsupervised method, i.e. how is recognized that two images have been closely aligned?"

### *2.2. Searching for related studies*

In the second step, we focus on determining appropriate literature, as strategies by the research question and purpose. To ensure that any similar systematic reviews have not been conducted, a search on Cochrane Library, PubMed, Scopus, and Science Direct using the selected keywords for the articles between January 2013 and December 2020 was made. Two authors examined all studies to identify the most relevant articles, as determined in the next stage.

### *2.3. Identifying relevant studies*

This study was conducted with common terms in image registration. To reach more speciality, i.e. to decrease non-related studies, we searched with synonymous words as well as the "AND" operator. The words, including title, abstract, and keyword were combined to find the related articles. The "control mesh" keyword in PubMed database was also used to find words, which were related to the article (Table 1). Here, unlike from the linearity we have in systematic reviews, the procedure of study selection this review was iterative and therefore more challenging to capture.

Articles before January 2013 (as there were no related articles before this year), articles with English abstracts, published in languages other than English, and reports or letters were excluded.

### *2.4. Synthesizing and tabulating key information*

Investigated all articles, some data were extracted and tabulated in forms, including, the region of interest (ROIs), datasets, techniques, similarity metrics, modalities, comparison methods, evaluation metrics, time, and novelties. All forms were synthesized by item and the findings were shown in frequency descriptive tables and graphs using Excel software (version 2016) [57]. While above procedure is known as data extraction within the topic of a systematic review and would be explained per specific statistical methods within the standard process of mapping or tabulating the data for a scoping review, we described data tabulating and mapping according to Arksey and O'Malley [55].

Table 1: Search strategy for unsupervised medical image registration.

| **Strategy** | **1 AND 2 AND 3** |
|---|---|
| #1 | (“Deep learning” [TIAB] OR “unsupervised deep learning”[TIAB] OR “unsupervised deep learning method” [TIAB] OR “unsupervised deep learning approach”[TIAB]) |
| #2 | (“convolutional neural networks” [TIAB] OR “deep convolutional neural networks” [TIAB] OR “unsupervised neural networks” [TIAB] OR “unsupervised convolutional neural networks” [TIAB]) |
| #3 | ("medical image registration")[Title]) OR "image registration"[Title]) OR "registration"[Title]) OR "unsupervised image registration"[Title]) OR "supervised image registration"[Title]) |

### *2.5. Summarizing and reporting the results*

In the last step, we went through the process of building an analytic workflow that explains the breadth of the state of art and determines priority fields within the literatures. Here, in our searched databases, 525 articles were selected; however, 185 articles were excluded due to the overlap in database. By scrutinizing the remaining 340 titles, 148 papers were excluded because of incompatibility with the research title. The remaining 192 articles were also reviewed based on abstracts. Consequently, 137 abstracts were excluded due to the lack of adequate compliance with the research objective. Finally, 55 articles were selected to study the full texts (See Fig. 1).

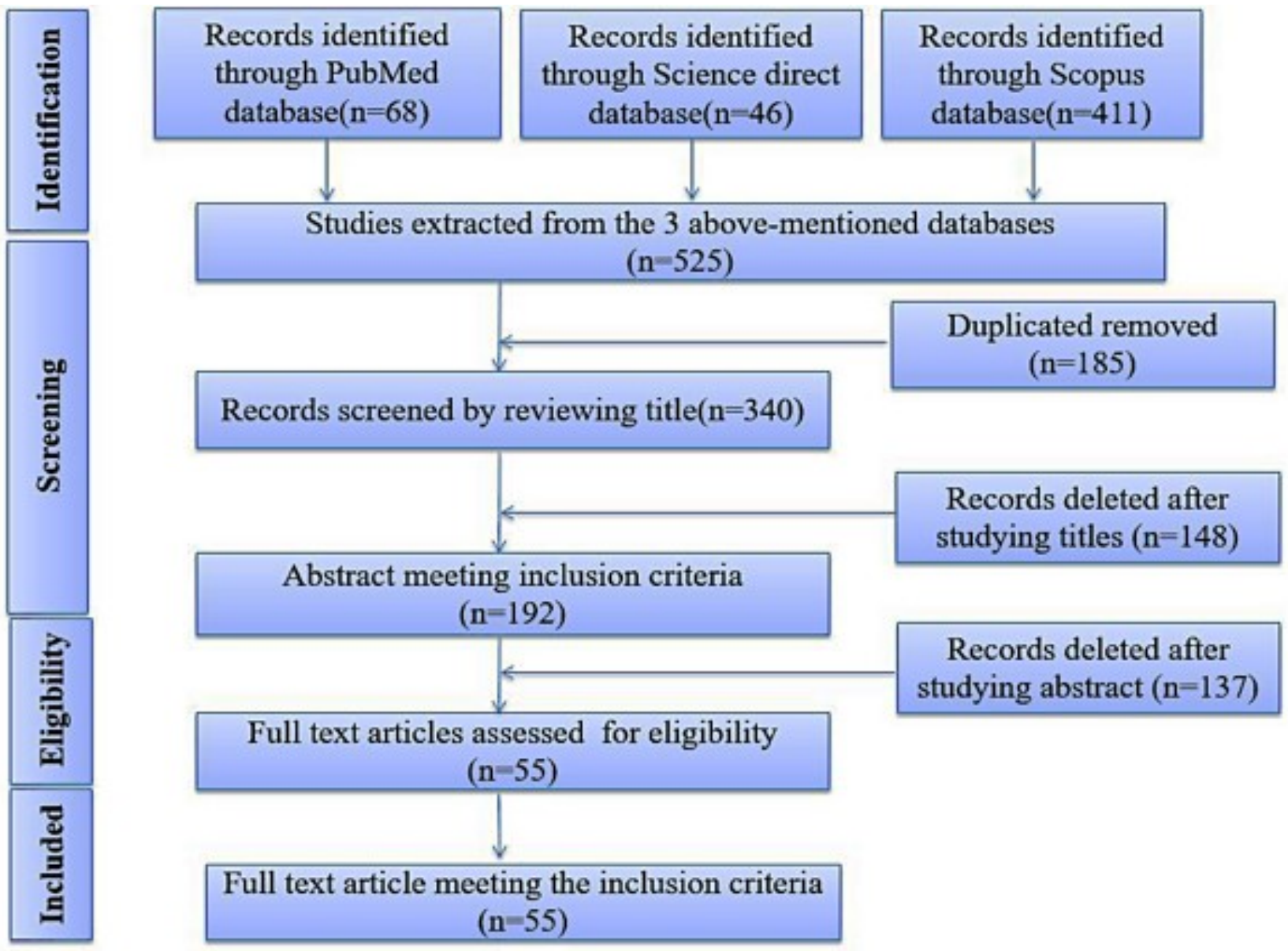

Figure 1: **PRISMA flow diagram for the scoping review process.**

## 3. The method of unsupervised transformation

In general, from training process viewpoint, deep learning-based medical image registration methods can be classified into two main classes, supervised, and unsupervised methods [58, 59]. In the supervised learning, neural network learns to do a certain work by minimizing a predefined loss function via optimization [2]. This refers to how the training databases are prepared and how the networks are trained using the databases. Here, the learning procedure of a network is performed according to the desired output which is already known in the training databases [60]. On the other hand, in unsupervised learning we don't have any target output available in the trained database. This means the desired target transformation parameters are absent in the trained databases [2]. In better words, unsupervised learning also refers to self-supervised learning due to the warped image is constructed from one of the input image pair and compared to another input image for
supervision.

Most of the researchers strongly focused on unsupervised image registration, trained without any need for a ground-truth [61, 60]. In this technique, data augmentation methods are used more in preference to a big ground-truth set. As seen in Fig. 2, both images, including fixed and moving images as a matrix with two channels, are firstly inputted into the localization network, mostly a CNN; secondly, this network estimates the transformation parameters for the affine or deformable cases. Next, a module is needed to exert the transform parameters on the moving image using a resampler, which is differentiable, to obtain the warped image [62]. It is worth mentioning that to train the CNN and obtain the best loss function, based on gradient descent backpropagation, the network (CNN) should be able to obtain the differentiable parameters. Thus, the differentiable function called spatial transformer network (STN) has been recently used, consisting of CNN and resampler [63]. Then fixed and warped images are compared based on the similarity metrics such as MI in multimodal images and mean square error (MSE) in unimodal images. Finally, the loss function is computed based on the similarity metrics and the network is trained by regularizer, smoothing transformation parameter until the best loss function is obtained by enough iteration [64, 65].

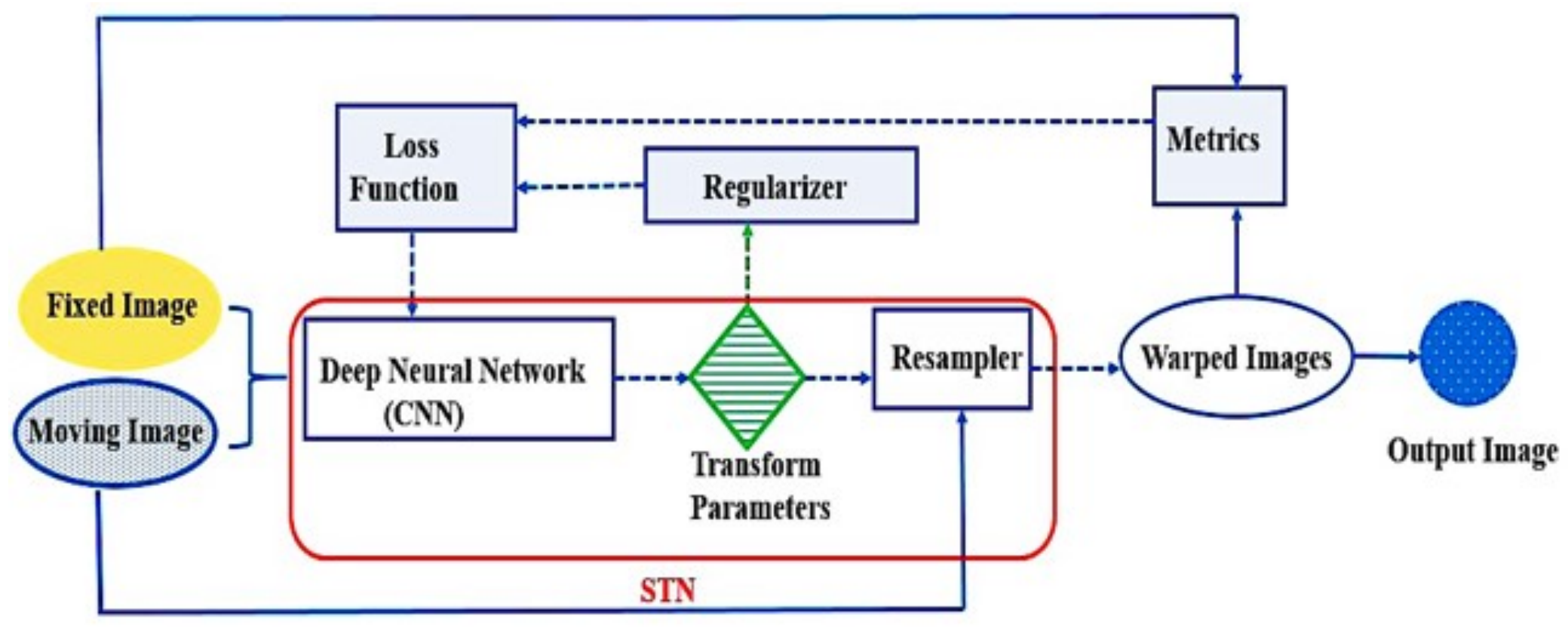

Figure 2: **The framework of unsupervised medical image registration..**

## 4. Literature review analysis

In image registration, both multimodal (the fixed and moving images have different modalities such as MRI and CT) and unimodal image registration have been introduced in the litratures. The establishment of multimodal has been more challenging and difficult than unimodal image registration due to different modalities, and mechanical similarity measures such as mean square error (MSE) and local cross-correlation (LCC). In fact, these metrics cannot be used in multimodal image registration and instead some other similarity measures such as MI and SIFT are used which are more inefficient and complicated. Moreover, the network trained by these similarity measures acquire this inefficiency as well. Thus, unimodal image registration is more successful than multimodal registration [60]. As some examples from the comparison of two methods, Wu et al. used hybrid Independent Subspace Analysis as well as CNN for the unimodal images registration (MRI) for automatically extracting and feeding features to Hammer for final deformable registration [66]. Wu et al. also conducted a study using Convolutional Stacked Autoencoder (SAE) on unimodal mages (MRI ) to extract and feed features for the deformable registration [67], which can be good evidence of the above argument as SAE was used instead of conventional multimodal measures such as MI.

Most of studies conducted on image registration have been in unimodal image registration such as Balakrishnan et al. [68], Stergios et al. [69], Shu et al. [70], Krebs et al. [71], Kearney et al. [72] using CNN; however, Ferrante et al. did a study by transfer-learning in CNN for zero-shot multimodal DIR in 2018 [73]. Besides, Zhang et al. [74], Yu et al. [75], Che et al. [76], and Theljani and Chen [77] conducted studies on multimodal image registration using CNN from 2010-2018. To the best of our knowledge, there are few studies on multimodal image registration from 2013 to 2020. However, due to a variety of CPUs and GPUs, datasets, techniques, ROIs, and similarity metrics, it would not be well advised to compare the results of unimodal and multimodal approaches to select the better one. According to the number of studies conducted in the literature, we can just report those registration methods that have been more successful in unimodal registration than multimodal registration. Moreover, all experimental results show that models with single-mode images than other models can reach the effective image registration performance with high accuracy as well as computational speed. As seen in Fig. 3, the unimodal and multimodal studies contain about 82% and 18% of all studies, respectively.

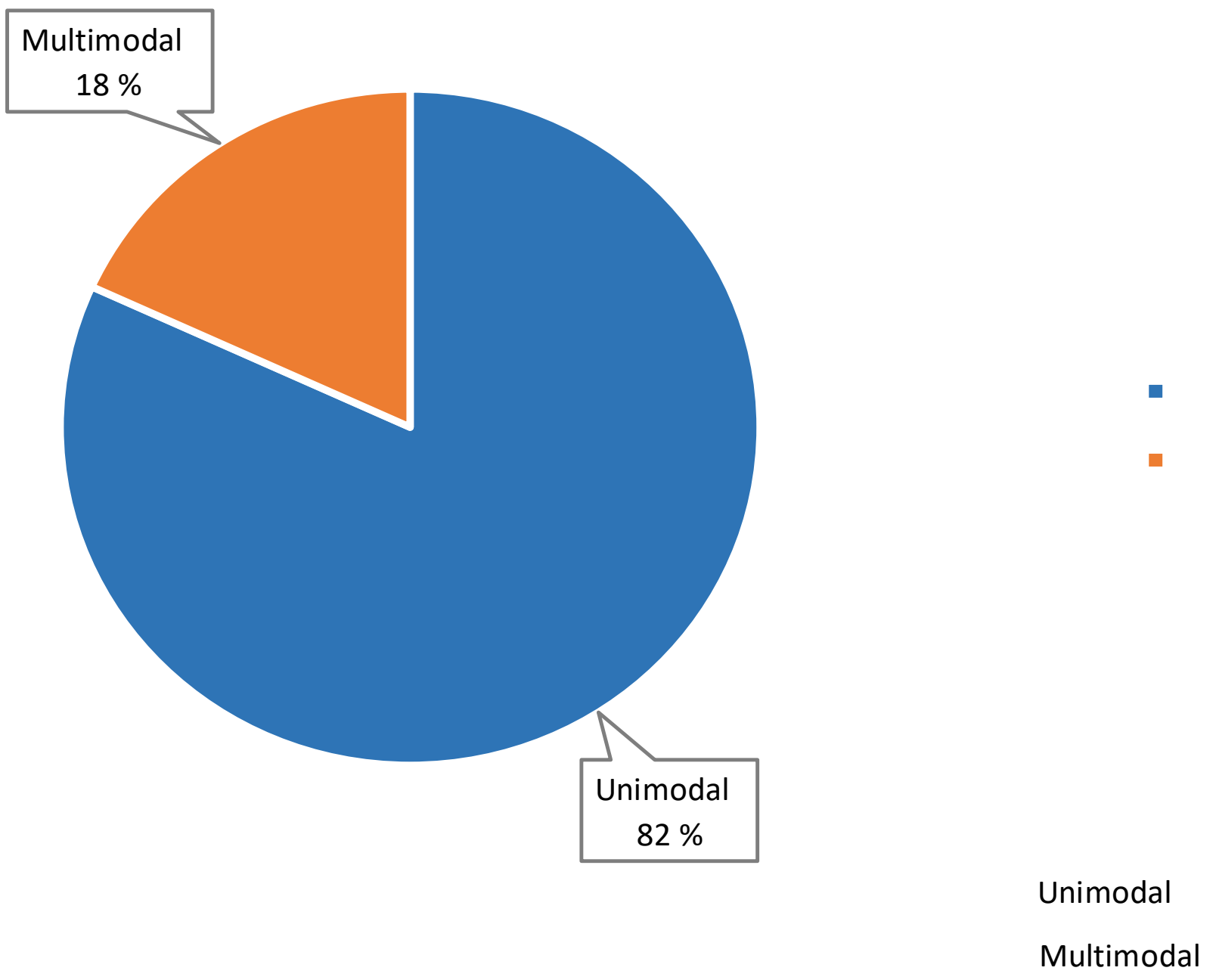

Figure 3: **Unimodal versus multimodal registration versus the number of publications.**

To simplify and improve future study on medical image registration using deep learning, we have summarised some of the publicly available databases used for developing the registration method in Fig. 4. For more infomation regarding the datasets see Ref. [78]. The more common datasets [3, 64, 68, 79] utilized for implementing and evaluating studies are revealed in this figure. The most frequent datasets used are as follows: 1) Private, 2) ADNI, 3) LPBA40 datasets. Moreover, the most frequent ROIs according to the publication numbers are also shown in Fig. 5 that the brain and cardiac are the most frequent organs, respectively, and head and neck and whole-body are studied less. The reasons for choosing the brain as the most frequent organ can be that it is surrounded by the skull, leading to simpler and faster alignment procedure, and also the interest structures are often more visible in different modalities such as MRI and CT, resulting in the more precise

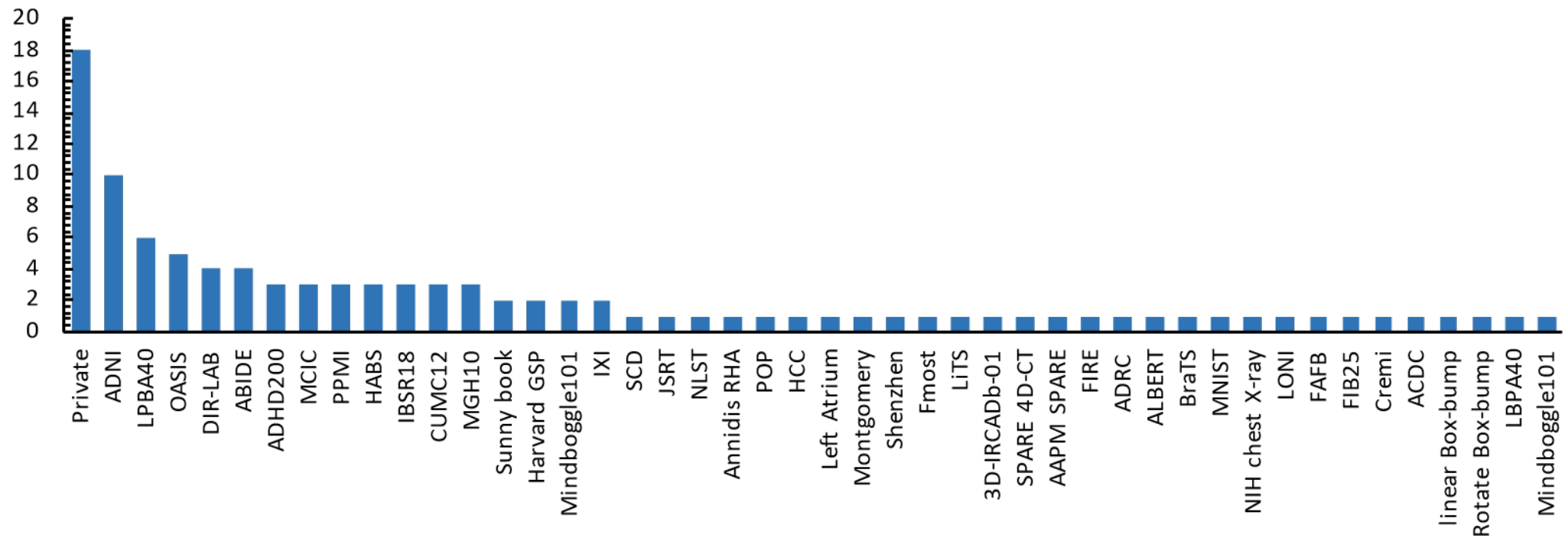

Figure 4: **The datasets versus the number of publications.**

and accurate validation procedure.

Fig. 6 shows the pie diagram of the most frequent modalities utilized in the literatures. MRI [3, 80, 81, 64, 66, 68, 69, 71, 73, 79, 82, 83, 84, 85] and CT [86, 87, 88] imaging were used at about 61% and 22%, respectively. In the next rank, X-ray [73, 89, 90] constitutes about 5%, and there are Multispectral imaging (MSI) [91, 92], the fluorescein angiographic image [93, 94], and the optical coherence tomography [94] fundus image for retina imaging, accumulated about 8%.

Based on similarity metrics, a loss function optimizes the values of the transformation parameter in neural network methods, i.e. similarity metrics are used to train the neural networks. Fig. 7 shows a bar chart of similarity metrics that the most frequent of them were normalized cross-correlation (NCC), cross-correlation (CC), and mean square error (MSE), respectively. In each evaluated study, the result of the proposed method was compared with the results of other methods, including conventional or learning-based methods. The most frequent method was Symmetric Normalization (Syn),

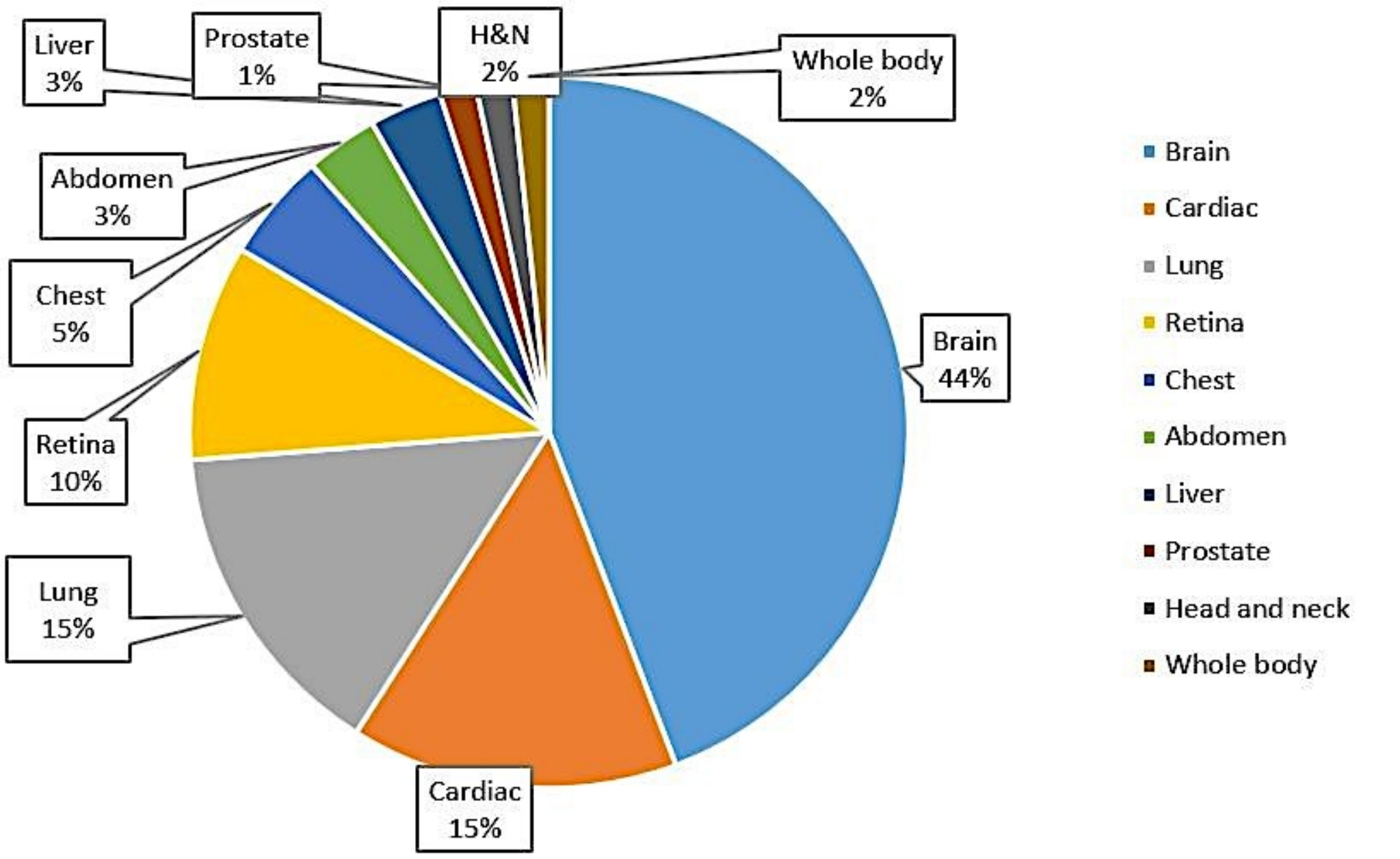

Figure 5: **The organs of interest versus the number of publications.**

as seen in Fig. 8.

By training the network and obtaining the results, there is a need to evaluate metrics for assessing and comparing the results with the results of the other studies. The most common evaluation metrics is the Dice similarity coefficient (DSC). The information about these metrics is seen in Fig. 9. Note, some of the evaluation metrics are almost same as similarity metrics such as MI, MSE, but it would be better not use them as the same as it that may make the invalid results and decreasing the accuracy [60].

Since DSC shows how much two inputted images overlap together and is also the most frequent evaluation metric in all assesses studies, it plays a key role in image registration. Hence, in this study, we reported mean DSCs

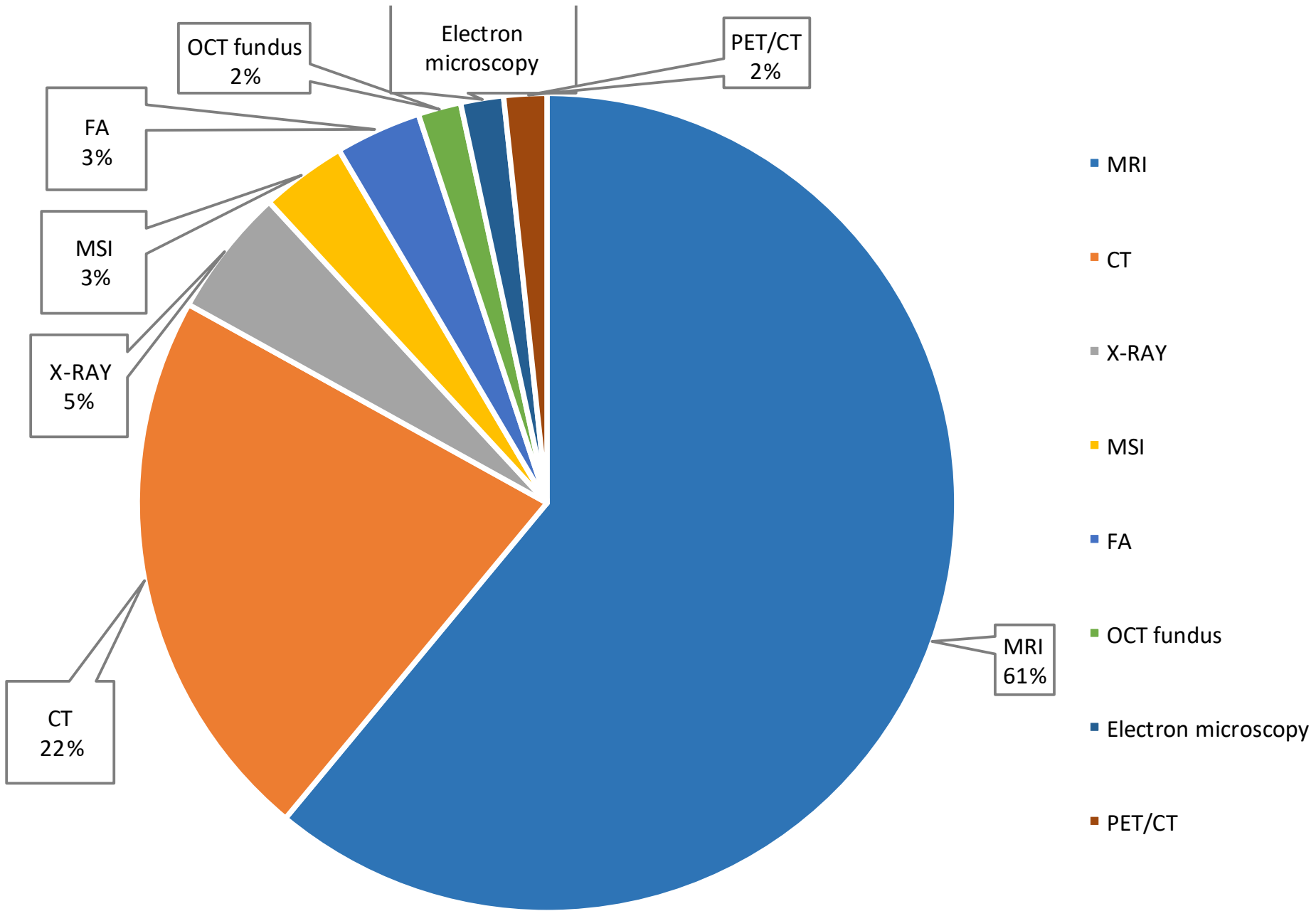

Figure 6: **Modalities used in the articles.**

obtained in all the reviewed papers for different techniques, modalities, and ROIs. Furthermore, according to Fig. 9, the second evaluation metric is the computing ”time” that some of the studies reported it based on using either GPU or CPU. Also some others just reported it with “time” and the rest do not report anything. Besides, due to the importance of the brain organ, as seen in Fig. 5, than the other organs, mean DSCs and time of brain studies, which are just obtained based on MRI modality, were separately reported in Table 2 and other organs are presented in Table 3. From neural network architecture viewpoint, in addition to CNN, some other networks are also reported such as Fast Image Registration (FAIM), Inverse-Consistent Deep Network (ICNet), Convolutional Stacked Autoencoder (SAE), the Structure-

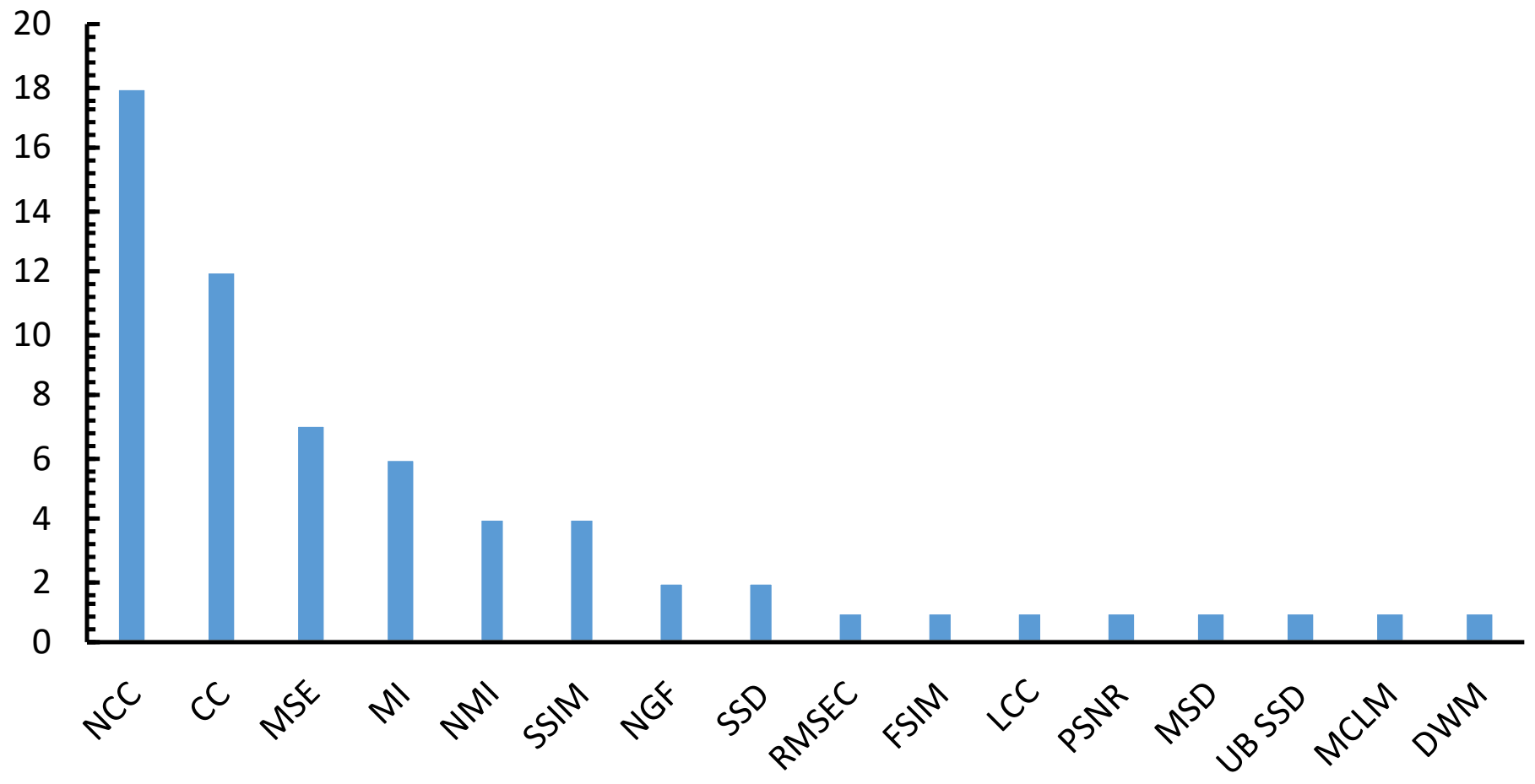

Figure 7: **The similarity metrics for the approaches versus the number of publications.**

Driven Regression Network (SDRN), Deep Groupwise Registration Framework (DGR-Net), Conditional Variational Autoencoder (CVAE) and Cycle Image Registration Network (CIRNet).

Most of the studies computed mean DSCs for different datasets and sometimes various parts of brain tissues such as white matter (WM), gray matter (GM), Cerebrospinal fluid (CSF), and Ventricle (V). Thus, we reported the best results obtained in each study in Table 2. Furthermore, some reported results need more explanation due to the complexity of the study such as the variety of datasets, subjects, and dimensions (2D-3D-4D). In short, Cheng, et al. conducted a study based on CNN on two datasets, including LPBA40 and IBSR18, and mean DSCs were revealed for WM, GM, and CSF that IBSR18 has the better mean DSCs in GM and WM than LPBA40, and also

CSF is reported 59.2 and 55.2 in LPBA40 and IBSR18, respectively [99]. A 3D-DNN model called U-ResNet, which is a kind of CNN, was designed and

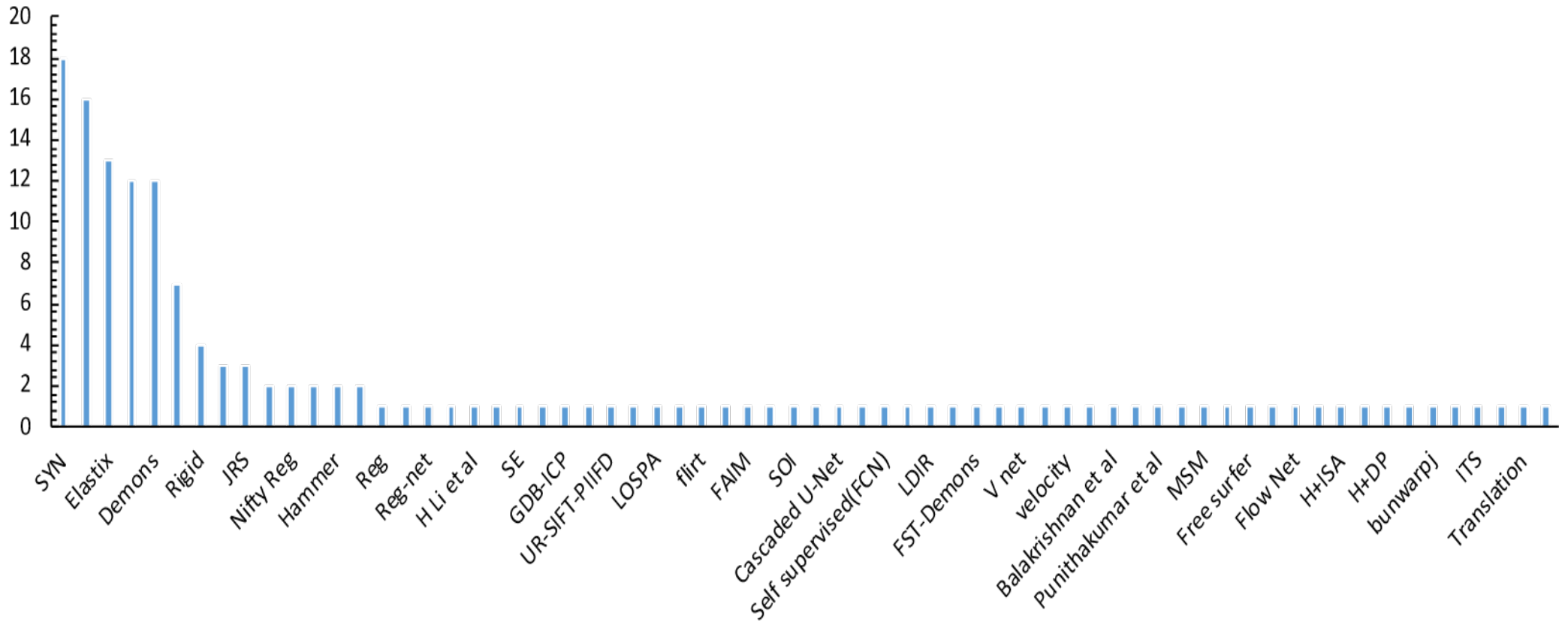

Figure 8: **Methods used for comparing the results versus the number of publications.**

implemented on OASIS 3 dataset for different categories of brain tissue such as brain stem (BS), caudate (Ca), CSF, amygdala (Am), cerebellum white matter (CblmWM) and cerebral white matter (CeblWM). Due to the categories used in this study, just CSF, V, and CeblWM as a WM are reported [98]. Zhang et al. conducted a study on 24 infant subjects and mean DSCs were obtained in WM, GM, and CSF for 0-, 3-, 6-, 9- and 12-month-old infants in two models [74] that the best results were almost obtained based on the second model and 12-month-old infants. Moreover, Hu et al. carried out a study on the brain-based new method of CNN, called Dual-Stream Pyramid

Registration Network (referred to as Dual-PRNet), and obtained mean DSCs 60.2, 69.0, 55.0, 69.5, and 61.8 for regions frontal, parietal, occipital, temporal and cingulate, respectively, on Mindboggle101 dataset [100]. Also, Liu et al. obtained mean DSCs on two datasets, including LPBA40 and MindBoggle101. As the mean DSCs in LPBA40 were higher, they were reported 71.1, 66.1, 66.0, 67.9, 69.1, 71.1, and 70.1 for frontal, parietal, occipital, temporal, cingulate, putamen, and hippo, respectively [101].

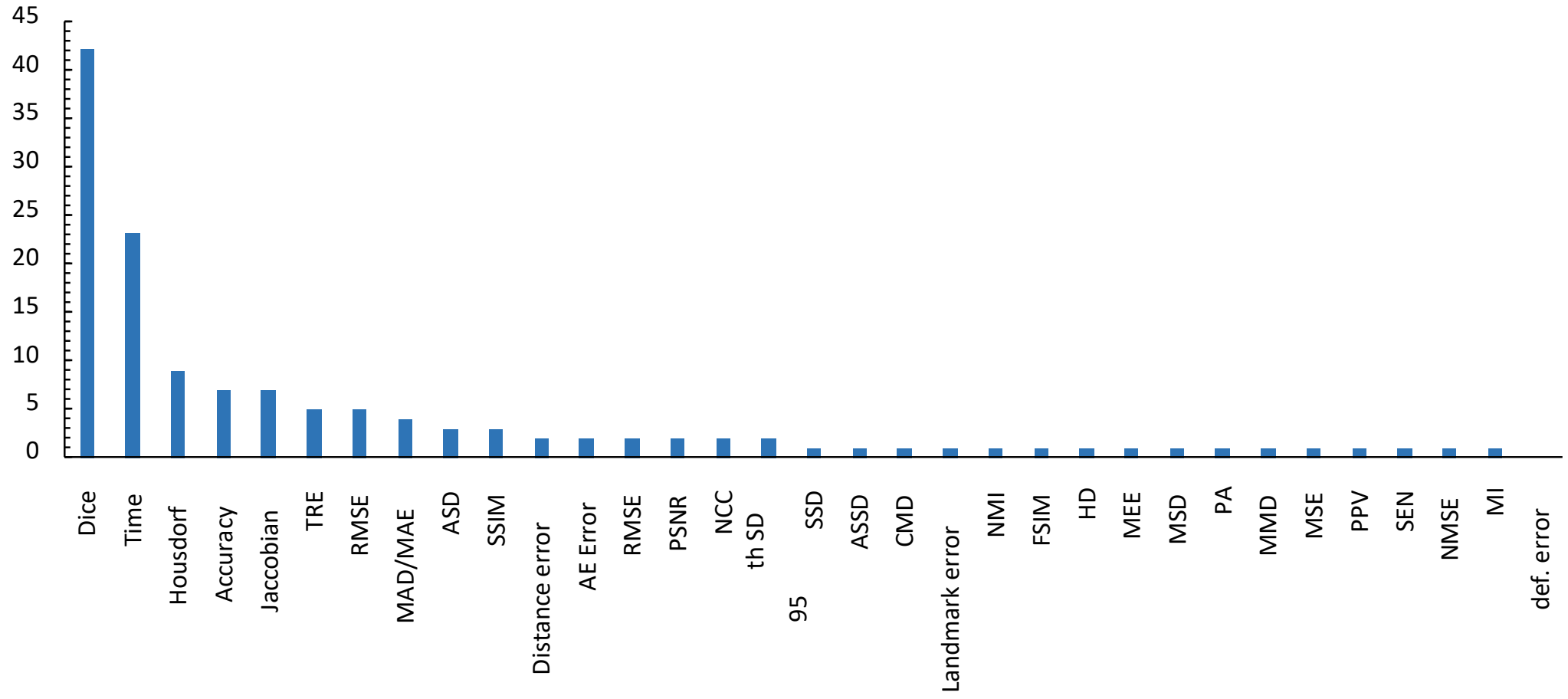

Figure 9: **The evaluation metrics for the approaches versus the publication numbers.**

Due to the importance of the unsupervised deep-learning technique as a state-of-the-art approach, many novelties have been presented in all studies. Thus, all novelties presented and introduced in every paper are summarized in Table 4 to help the active readers reach a better understanding of this field and employ these new ideas in future studies.

| Ref. | Novelty | Ref. | Novelty |
|---|---|---|---|
| [3] | 1. 3D global deformation field in a shot. 2. The use of FCN as a new multi-scale framework. 3. Training based on image similarity. | [32] | 1. An unsupervised learningbased method in registering deformable image of thoracic CT scans. |
| [80] | 1. No need for training data and rough pre-alignment of images. 2. Image registration in one pass. | [64] | 1. A probabilistic generative model and a learning-based inference algorithm. |
| [51] | 1. Introducing Voxel Morph, a speedy learning-based method for deformable, pairwise medical image registration. | [66] | 1. A stacked convolutional ISA network. |
| [65] | 1. A probabilistic technique for diffeomorphic image registration. 2. Registration for 3D images. 3. Novel scaling as well as squaring differentiable layers. | [68] | 1. No need to training data. 2. Different cost functions for different tasks. |
| [69] | 1. Combination of linear as well as deformable registration within a single optimization architecture and a coupled framework. 2. A modular and parameter-free implementation, independent of the various similarity metrics. | [70] | 1. The fast registration for microscopic images of neuron tissues. |
| [71] | 1. The use of a user-adjustable smoothness layer as well as differentiable exponentiation. 2. The first experiment on disease clustering as well as deformation transport. | [72] | 1. A deep unsupervised learning strategy for cone-beam CT (CBCT) to CT DIR. 2. The use of a deep convolu-tional inverse graphics network (DCIGN) |
| [73] | 1. The use of a simple one-shot learning strategy. | [74] | 1. Just one deformation field in spite of the other techniques. 2. The multi-contrast MRI as well as artificial images like gradient images with multiple input channels. 3. Training in an auto-context and the multi-scale loss function. |
| [76] | 1. A novel end-to-end network according to ResNet as well as U-Net as Encoder-decoder. | [77] | 1. The unsupervised deep learning technique for monoand multi-modal image regis- tration. |
| [75] | 1. Automatically extracting features among the moving PET images as well as fixed CT images. 2. Balancing the image similarity as well as anatomical structure deformation by the new cost function. | [79] | 1. A learned bicubic CatmullRom spline resampler for minimizing error in DVF resampling. 2. Learning complicated deformations as well as computing accurate, smooth, and reasonable DVFs. |
| [84] | 1. Performing Rigid as well as deformable registrations for PMR3 in addition to pCT4 images. 2. Mutual-information (MI)-based Cycle-GAN. 3. The use of UR-Net. | [82] | 1. The introduction of adjustable ConvNets designs for deformable and affine image registration. |
| [85] | 1. A new technique on FCNs. 2. The optimization of registration algorithm. | [86] | 1. VoxelMorph. 2. 3D deformable registration. |
| [87] | 1. A theoretical and direct coarse-to-fine registration transference from conventional as well as iterative registration schemes. | [88] | 1. Generating aligned pairs to train the discriminator network. 2. PatchGAN as a local quality measure in aligning images. |
| [89] | 1. The first anatomically constrained neural network for the image registration problem. | [90] | 1. Registration as well as segmentation in a single DL framework 2. Obtaining registration and segmentation by each other. |
| [91] | 1. The first technique for unsupervised deformable group-wise registration in mono/multi-modal MSI data for the same and various wavelengths, respectively. 2. The use of an iteratively updated representative template image according to PCA1. | [92] | 2. A new structure-driven regression framework trained based on the strategy of multiscale deformation fields. |
| [102] | 1. The use of GANs. | [94] | 1. Feature-based method based on CNN for multimodal retinal image registration. 2. The use of deep step patterns (DeepSPa) algorithm. |
| [81] | 1. Registration for 3D images. 2. A non-rigid imageregistration algorithm with FCN. | [105] | 1. 3D CNNs. 2. Generating the deformation field of serial electron microscopy (EM ) images through the testing phase. |
| [95] | 1. A framework for the registration of spherical surface. 2. Solution of a Maximum a Posteriori (MAP) 3. The various features for functional alignment. | [96] | 1. Simultaneously learning as well as using the populationlevel statistics of the spatial transformations for regularizing the neural networks. |
| [97] | 1. No need for the training dataset. 2. Automatic estimation for important points. | [98] | 1. The first joint DR framework for simultaneous optimization of image segmentation as well as registration. |
| [106] | 1. Using an Inverse-Consistent deep Network (ICNet). 2. Using an anti-folding restriction. | [99] | 1. A novel cascaded 3D U-Net by dilated convolution for volumetric image registration. |
| [100] | 1. Developing a dual-stream encoder-decoder network. 2. Proposing a pyramid registration technique directly for estimating registration fields from convolutional features. | [101] | 1. A feature-level probabilistic technique |
| [103] | 1. The use of an unsupervised network with low-dimensional vectors as input. 2. Creating the fields of spatial transformation using embedding. | [107] | 1. DIR technique for 4D-CT lung images. 2. Fast predict DVF in a few forward predictions. 3. An unsupervised deep learning method for 4D-CT lung DIR. |
| [108] | 1. Tracing changes in the retina structure across modalities by vessel segmentation as well as automatic landmark detection. | [109] | 1. The introduction of DeepMapi technique. |
| [110] | 1. A Multi-scale DIR framework with unsupervised Joint training of Convolutional Neural Network (MJ-CNN). | [111] | 1. The first multi-scale unsupervised deep-learning-based on DIR for abdominal 4D-CT images. 2. Being integrated into MS-DIRNet. |
| [112] | 1. The use of CNN with cycle-consistency for registration. 2. An improvement to topology preservation. 3. Training the proposed method by a diverse source as well as target images in multiphase contrast-enhanced CT (CECT) acquisition. | [113] | 1. Deep learning technique for registering deformable medical images. 2. Learning posterior distribution over the weights using backpropagation. |

Table 4: Novelties presented in the literature reviewed.

## 5. Discussion

Recent years have shown a highly increase in the number of published studies on using deep learning-based in image registration. In fact, by releasing several open-source deep learning software libraries such as Tensorflow, Keras and Pytorch the development of deep learning-based studies experienced a huge rise after 2015. In spite of it is a few years since deep learningbased approaches were used for medical image registration, the use of them has seen different changes. The evolution of deep learning-based approaches is characterised into four stages. The best reference for explanation of these stages can be found in [78]. Briefly, the first stage tried to use deep learning networks for extracting features, which were utilized to guide traditional registration methods. Next stage worked on addressing a key limitation of iterative traditional registration methods. The purpose of several supervised networks introduced in this stage was mainly to make the process of registration fast. Model training here on appropriate image pairs are many-folds faster than conventional iterative registration methods during testing. However, supervised approaches need ground-truth for spatial transformations for training step, which are elaborating task to obtain in most clinical applications [114]. for this reason, unsupervised methods were presented. These methods showed comparable registration accuracy and speed with supervised ones, while they don't need labels at all.

The current scoping review on the medical image registration literature based on unsupervised deep neural networks shows that the proposed methods aimed at improving two parameters as evaluation metrics, including the mean DSCs and registration running-time, for greater performance, i.e. it is expected that to extremely improve performance a network, it is not overengineered due to the incremental number of layers, parameters, and connections. In most studies, the authors have claimed that their proposed techniques can register two inputted images in less possible time and with the highest performance, both effective and efficient for clinical application [79, 47]. The proposed review aims to just report some data such as modalities, datasets, ROIs, techniques, similarity metrics in different analysis and to cover evaluation metrics, including mean DSCs and time. Since there are the variety of CPUs and GPUs, datasets, techniques, ROIs, and similarity metrics, it would not be well advised to compare the results of unimodal and multimodal approaches to select the better one. Hence, we just report their findings. Moreover, as the unsupervised image registration method does not require any experts to annotate the ground-truths, many researchers have recently been engrossed in this approach. The desired results reveal that this method has been more successful on unimodal registration than multimodal registration. On the other hand, in some clinical applications, multimodal registration is needed due to valuable information in different kinds of images such as CT and MRI. Hence there is a need to do more research on unsupervised image registration methods to address different kind of challenging problems [52]. To the best of our knowledge, some review studies have been conducted on image registration [2, 78], there has been no comprehensive scoping review on unsupervised technique, reporting findings of all studies intricately in order to help those active readers, who are riveted by this field, achieve deep insight into this exciting field and can compare the results comprehensively in some tables, presented in this study.

## 6. Conclusions

In this review, we conducted a comprehensive scoping review on medical image registration studies using unsupervised deep learning for articles published over the past 8 years. The findings were extracted and categorized according to the datasets, similarity metrics, modalities, transformation, and ROIs. Additionally, evaluation metrics, including DSCs and time were reported in tables; it was found that the least run-time was 0.0027s and 0.021s, respectively, for brain and retinal images [92, 94], and the best mean DSC was 97% for the brain images [79]. Furthermore, the novelties of all papers are summarized (Table 4) to provide new ideas for active readers in future studies. Based on the recent success, unsupervised deep learning has an enormous impact on medical image registration.

At the end, although deep learning-based approaches have addressed different challenges in medical image registration and achieved faster and more accurate results than traditional ones, there are several topics that are good to be tackled in this area in future. These challenges includes: lack of appropriate publicly available database which has limited the development of deep learning-based approaches. Second is related to clinical applications of deep learning-based methods which are the final goal for all the medical image processing and analysis approaches. Until now, in spite of various well-proved deep learning-based image registration we don't see any of these methods deployed in a clinical setting. Finally, by considering from the statistics of the published studies, both supervised and unsupervised prediction methods are sort of equally worked with a small difference number of publications in either category. Either supervised or unsupervised approaches have their own pros and cons. We believe that more research will be concentrated on combination of supervised and unsupervised approaches in the future.

Table 2: All mean DSCs and time for brain in MRI modality. W: White matter, G: Gray matter, CSF: Cerebrospinal fluid, V: Ventricle.

| Ref. | Technique | Mean DSCs (%) | GPU(s) | CPU(s) | Time(s) |
|---|---|---|---|---|---|
| [85] | FCN | 72 | – | – | 0.2 |
| [68] | CNN | 75 | 0.55 | 144 | – |
| [64] | CNN | 75.3 | 0.45 | 51 | – |
| [62] | CNN | 75.3 | 0.45 | 57 | – |
| [65] | CNN | 75.4 | 0.47 | 84.2 | – |
| [90] | CNN | 91.1 | 1.9 | 87.7 | – |
| [95] | CNN | 88.2 | – | – | 44.4 |
| [70] | CNN | 89.51 | – | – | 0.19 |
| [96] | CNN | 98 | – | – | 0.02 |
| [97] | CNN | 91 | – | – | 0.2 |
| [79] | CNN | 97 | <1 | 2.94 | – |
| [3] | FCN | $84.83^{G}, 80.30^{W}$ | 0.74 | – | – |
| [98] | CNN | $71.4^{W}, 44.6^{C}, 70.4^{V}$ | – | – | – |
| [74] | CNN | $78.7^{G}, 77.0^{W}, 59.40^{C}$ | – | – | – |
| [99] | ICNet | $80.6^{G}, 80.6^{W}, 83.6^{C}$ | – | – | – |
| [100] | CNN | $77.6^{G}, 82.6^{W}, 55.2^{C}$ | – | – | 0.003 |
| [67] | SAE | $78.6^{G}, 88.1^{W}, 93.0^{V}$ | – | – | – |
| [66] | ISA | $78.6^{G}$,$88.1^{W}$,$95.0^{V}$ | – | – | – |

Table 3: All mean DSCs for different organs except brain. Cardiac: ventricular endocardium (LVE), right ventricular endocardium (RVE), left ventricular mass (LVM) Retina: Optic (O), Vascular (V).

| **Ref.** | **ROI** | **Model** | **Modality** | **Mean DSCs(%)** | **GPU(s)** | **CPU(s)** | **Time(s)** |
|---|---|---|---|---|---|---|---|
| [69] | Lung | CNN | MRI | 91.48 | 0.5 | – | – |
| [79] | Lung | CNN | CT | 0.90 | – | – | 2.94 |
| [82] | Lung | CNN | MRI, X-ray | 77 | – | 20.36 | – |
| [87] | Lung | CNN | CT | 92.1 | – | – | – |
| [90] | Lung | GAN | X-ray | 88.9 | – | – | 0.5 |
| [32] | Lung | CNN | CT | 96 | – | – | 0.17 |
| [84] | Liver | GAN | MRI, CT | 90.59 | – | – | – |
| [91] | Retina | CNN | MSI | $96.7^{O}$, $73.2^{V}$ | 0.5 | – | – |
| [92] | Retina | SDRN | fundus imaging | 75.3 | – | – | 0.02 |
| [102] | Retina | GAN | fundus imaging | 94.6 | – | – | 0.8 |
| [76] | Retina | DGRNet | MSI | $97^{O}$, $75.2^{V}$ | – | – | – |
| [71] | Cardiac | CVAE | MRI | 78.3 | 0.32 | 108 | – |
| [80] | Cardiac | CNN | MRI | 80 | – | 0.5 | – |
| [83] | Cardiac | CVAE | MRI | 81.2 | – | – | 0.32 |
| [94] | Cardiac | GAN | MRI | 85 | – | – | 0.7, 0.8 |
| [103] | Cardiac | FCN | MRI | 89 | – | – | – |
| [41] | Cardiac | CNN | MRI | 89 | 0.04 | 2.35 | – |
| [104] | Cardiac | CIRNet | MRI, CT | $83.5^{LVE}$, $80.8_{LVE}$, $72.6_{RVE}$ | 0.245 | 0.245 | – |